\documentclass[aps,prl,twocolumn,groupedaddress,showpacs]{revtex4}
\usepackage{graphicx}
\usepackage{amssymb}

\begin{document}


\title{Dynamics of shallow impact cratering}


\author{M.A. Ambroso$^1$, R.D. Kamien$^2$, and D.J. Durian$^{1,2}$}
\affiliation{
     $^{1}$Department of Physics and Astronomy, University of
     California, Los Angeles, CA 90095-1547, USA\break
     $^{2}$Department of Physics and Astronomy, University of
     Pennsylvania, Philadelphia, PA 19104-6396, USA
}


\date{\today}

\begin{abstract}
We present data for the time-dependence of wooden spheres
penetrating into a loose non-cohesive packing of glass beads. The
stopping time is a factor of three longer than the time
$d/v_\circ$ needed to travel the total penetration distance $d$ at
the impact speed $v_\circ$. The acceleration decreases
monotonically throughout the impact. These kinematics are modelled
by a position- and velocity-dependent stopping force that is
constrained to reproduce prior observations for the scaling of the
penetration depth with the total drop distance.
\end{abstract}

\pacs{45.70.Ht, 45.70.Cc, 83.80.Fg, 89.75.Da}


\maketitle



\begin{figure}
\includegraphics[width=3.00in]{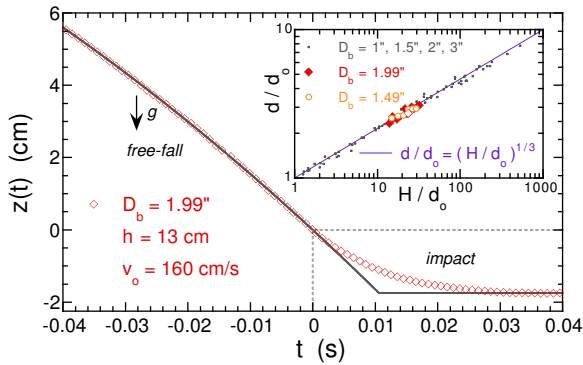}
\caption{The vertical position $z$ of the bottom of a sphere vs
time $t$, for typical conditions as labelled. The impact begins at
$\{z=0,t=0\}$, and gravity points in the $-z$ direction.  Only
every $1/100^{th}$ datum is shown. The inset shows the final
penetration depth vs total drop distance, both scaled by the
minimum penetration depth $d_\circ$. The small symbols represent
data for four wooden spheres from Ref.~\protect\cite{mike1}.  The
large symbols represent new runs, for which in this manuscript we
report on impact dynamics ($\diamond$ is for the example in the
main plot). The values of $d_\circ$ are 0.63~cm and 0.44~cm for
the 1.99 inch and 1.49 inch diameter spheres, respectively.
\label{Z06vsT}}
\end{figure}

Granular impact is a phenomenon of natural interest. One focus of
recent work is the size and morphology of the crater, and the
analogy with planetary cratering \cite{amato,hans,deBruyn}.
Another is the dramatic splash produced by the collapsing void
\cite{siggi,detlef2}. Still another is the final depth of
penetration, $d$, because it probes granular mechanics via the
depth-averaged stopping force: $\langle F_s\rangle = mgH/d$, where
$m$ is the projectile mass, $g=980$~cm/s$^2$, $h$ is the free-fall
distance, and $H=h+d$ is the total drop distance
\cite{jun,katie,deBruyn2,zheng,mike1}. For shallow impact by
spheres \cite{jun,mike1}, and for deeper impacts by cylinders with
various tip shapes \cite{katie}, the penetration depth scales as
\begin{equation}
    d/d_\circ=(H/d_\circ)^{1/3},
\label{junlaw}
\end{equation}
where $d_\circ$ is the minimum penetration for $h=0$.  The inset
of Fig.~\ref{Z06vsT} shows Eq.~(\ref{junlaw}) agreeing with data
over nearly three decades in $H$. While this constrains the
stopping force, it does not reveal a unique form.  For example,
Eq.~(\ref{junlaw}) is equally consistent with $F_s\propto z^2$ and
$F_s\propto v^{4/3}$, where $z$ and $v$ respectively are the
instantaneous depth and speed of the projectile \cite{jun}. Which,
if either, of these possible stopping forces is correct?  How does
the nature of the stopping force, and the resulting transfer of
energy from the projectile to the medium, conspire to produce the
subsequent granular splash and the final crater morphology?  The
unintuitive response of granular media to external forcing is a
topic of widespread interest beyond the specific example of
impact~\cite{brown, jnb, duran}.

Recently impact dynamics have been measured by high-speed video
\cite{ciamarra,detlef3,RPB04} and by an embedded accelerometer
\cite{PBU}.  In Ref.~\cite{ciamarra} the total upward force is
found to be $\Sigma F = -mg+(mg+kd)$.  The stopping force, in
parentheses, is independent of time but has a value that depends
on the impact speed $v_\circ$. Solution of $\Sigma F = ma$ gives a
penetration depth of $d=\sqrt{m{v_\circ}^2/(2k)}$.  In
Ref.~\cite{detlef3}, the total force is found to be $\Sigma F =
-mg+k|z|$. The stopping force is Coulomb friction and increases
with time. Solution of $\Sigma F = ma$ gives a penetration depth
of similar form to Eq.~(\ref{junlaw}):
$d/d_\circ=(H/d_\circ)^{1/2}$ with $d_\circ=2mg/k$.  In
Refs.~\cite{RPB04,PBU}, the acceleration decreases with time. The
various reported force laws thus appear contradictory, both in
terms of their time dependencies and in terms of their predicted
penetration depths. Furthermore, none of the reported force laws
is consistent with the $d\sim H^{1/3}$ observation of
Eq.~(\ref{junlaw}).

In this paper we measure cratering dynamics in the unexplored
regime of shallow impact, where the projectile never submerges.
Our approach is to measure position vs time with an optical
method, both faster and more precise than imaging. Like
Refs.~\cite{RPB04,PBU}, we find that the acceleration decreases
throughout impact.  Our theoretical approach is to consider
possible instantaneous force laws whose depth-averages reproduce
the observed scaling of Eq.~(\ref{junlaw}). The best candidate
depends on both position and speed, and suggests that the
seemingly disparate results of
Refs.~\cite{jun,ciamarra,detlef3,RPB04,PBU} may not be
contradictory but instead may represent limiting cases of a common
force law that holds for both shallow and deep impacts.

Our materials and penetration depth measurements are identical to
those of Ref.~\cite{mike1}.  The medium consists of glass beads,
diameter $D_g=0.30\pm0.05$~mm, prepared at 59\% packing fraction
by slowly turning off a fluidizing upflow of air. The projectiles
are wooden spheres of diameter $D_b=1.49$~inch or $D_b=1.99$~inch,
and of density $\rho_b=0.7$~g/cc.  These are held and dropped from
rest via a suction mechanism.  The drop distance and final
penetration depth are measured with a telescope mounted to a
height gauge.  Data for the wooden spheres of Ref.~\cite{mike1},
and for the dynamics runs reported here, are shown in the inset of
Fig.~\ref{Z06vsT} to obey Eq.~(\ref{junlaw}).

The time-dependent vertical position of the projectile is measured
optically. A laser ($\lambda=532$~nm) is placed about 1.5~m from
the sample, along with a cylindrical lens that fans the beam into
a thin sheet. An aperture is used to select the central portion of
the beam, where the intensity is nearly constant, and to set its
size to be slightly greater than $D_b$. On the other side of the
projectile, we align a second aperture equal to $D_b$. Behind this
we place a large planoconvex lens to focus the light onto a
photodiode. As such, the collected light intensity varies nearly
linearly with projectile position. In each run the photocurrent
starts at a maximum, decreases as the projectile freely falls onto
the beam, goes to zero when the beam is fully blocked, and then
increases as the projectile falls further and light passes over
its top; impact occurs during this last phase. To calibrate we
hold the ball in several known positions, with light passing above
and below, and we fit to the particular cubic polynomial expected
for an aligned Gaussian beam.

Typical depth vs time data are displayed in Fig.~\ref{Z06vsT}. The
impact occurs at $t=0$, and the projectile comes to rest in about
$0.03$~s.  The solid curve through the $t<0$ free-fall data is not
a fit, but rather $-v_\circ t-gt^2/2$ with $v_\circ=\sqrt{2gh}$.
We digitized the photocurrent using a 12-bit A/D converter
operating at $10^5$ points per second. Finer digitization and
faster capture could be achieved at marginally greater expense.
The fidelity with which impact dynamics can be captured by our
method exceeds high-speed video. Notwithstanding, there exist
certain limitations.  One is that for large enough drop heights
the grains splash into the laser beam. A lesser limitation is that
the minimum drop height was such that none of the beam was blocked
by the projectile prior to its release. Overall we achieved over a
factor of two variation in total drop distance for each sphere. To
differentiate position vs time data, we fit to a cubic polynomial
with Gaussian weighting that nearly vanishes at the edges. For
fitting windows that are too small, the velocity and acceleration
results are noisy; for fitting windows that are not too large, the
depth-averaged acceleration equals $gh/d$ as required by energy
conservation. This check gives confidence in both our data and our
differentiation procedures.


\begin{figure}
\includegraphics[width=3.00in]{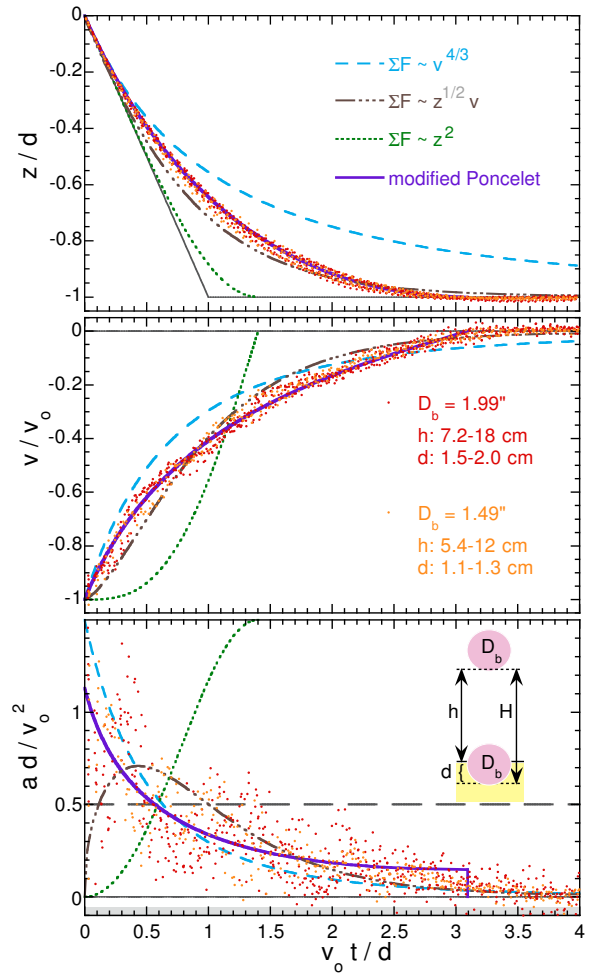}
\caption{Vertical components of position, velocity, and
acceleration vs time, for all runs; red (orange) dots are for
$D_b=1.99$ inch (1.49 inch). To within measurement errors, the
data collapse when lengths are scaled by $d$ and times are scaled
by $d/v_\circ$, where $d$ is the final penetration depth and
$v_\circ$ is the impact speed. By construction, the scaled
position data all decay from $0$ to $-1$ with an initial slope of
$-1$ (gray lines).  Note that the acceleration is not constant
(gray dashed line $a d/{v_\circ}^2=1/2$), but rather decreases
with time. The data rule out the power-law forces consistent with
Eq.~(\protect{\ref{junlaw}}), but are consistent with a modified
Poncelet model. \label{ZVAvsT}}
\end{figure}

Data for position-velocity-acceleration vs time, for both spheres
and all drop heights, are displayed in Fig.~\ref{ZVAvsT}. All
lengths are scaled by the final penetration depth $d$, and all
times are scaled by the time $d/v_\circ$ required to move a
distance $d$ at the impact speed $v_\circ$.  To within measurement
error, the data all collapse according to this scaling.  By
construction, the scaled position data must decay from $0$ to $-1$
with an initial slope of $-1$; the scaled velocity data must decay
from $-1$ to $0$; and the depth-average of the scaled acceleration
must equal $1/2$.  We find that the spheres all come to rest at
about $3v_\circ t/d$, roughly three times longer than if they
moved the same distance at constant speed.  Since $v_\circ\sim
h^{1/2}$ and $d\sim H^{1/3}$ have similar scaling, the impact
duration is nearly constant as in Refs.~\cite{ciamarra,PBU}.  We
also find that the acceleration decreases with time, in accord
with Refs.~\cite{RPB04,PBU} but in contrast to
Refs.~\cite{ciamarra,detlef3}. Note that the scaled value of
gravitational acceleration is $-gd/{v_\circ}^2=-d/(2h)$; this
ranges from $-0.05$ to $-0.10$ for our runs (shaded gray region in
Fig.~\ref{ZVAvsT}c) and is generally small compared to the
projectile acceleration.

Our impact dynamics data can now be compared with expectations for
various candidate force laws.  For example the simple {\it ad-hoc}
form $\Sigma F = -mg + k |z|^\alpha |v|^\beta$ best agrees with
Eq.~(\ref{junlaw}) if the exponents are related by $\beta =
(4-2\alpha)/3$.  The agreement becomes exact for $\Sigma F = -mg +
k |z|^2$. Predictions for this special case,
$\{\alpha=2,\beta=0\}$, and also for $\{\alpha=0,\beta=4/3\}$, are
displayed with the scaled data in Fig.~\ref{ZVAvsT}. Evidently,
the decay of $z(t)$ is too fast for the former and too slow for
the latter. The actual behavior lies between these extremes.  A
marginally-acceptable fit, also shown, is attained for
$\{\alpha=1/2,\beta=1\}$.

\begin{figure}
\includegraphics[width=3.00in]{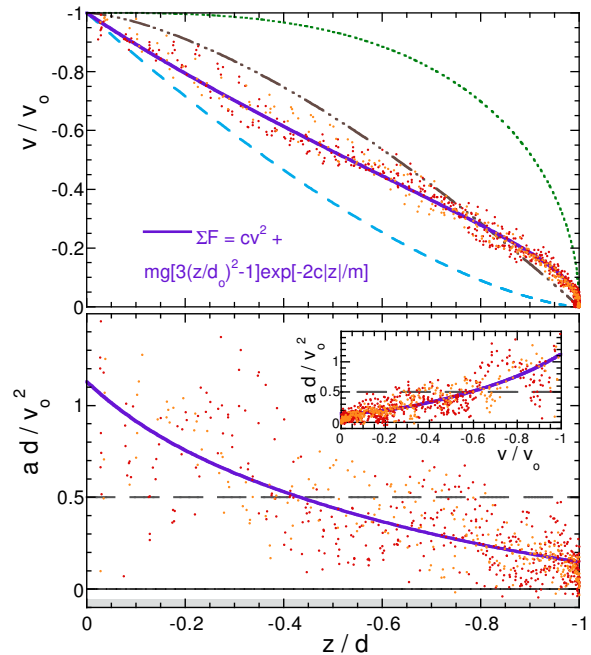}
\caption{Velocity and acceleration vs position, and acceleration
vs velocity (inset), for all runs, scaled as in
Fig.~\protect\ref{ZVAvsT}; red (orange) dots are for $D_b=1.99$
inch (1.49 inch).  Excellent fits to the modified Poncelet model
are shown as solid purple curves.  The various-dashed curves
represent the same power-law forces as in
Fig.~\protect\ref{ZVAvsT}.\label{VAvsZ}}
\end{figure}

Better fits can be achieved if the stopping force equals a
constant plus a term that grows with speed.  If the drag is
viscous, then the force law is given by the Bingham model, $\Sigma
F = -mg+(F_\circ+b|v|)$.  If the drag is inertial, then the force
law is given by the Poncelet model, $\Sigma F =
-mg+(F_\circ+cv^2)$. The Bingham model has recently been advocated
for granular impact \cite{deBruyn2}, while the Poncelet model has
long been used for ballistics applications \cite{backman78}. For
both, position vs speed can be found by writing $a=v{\rm d}v/{\rm
d}z$, separating variables, and integrating. All our dynamics data
are shown again in phase space plots of scaled velocity and
acceleration vs depth, as well as acceleration vs speed, in
Fig.~\ref{VAvsZ}. The best one-parameter fits to these models give
$(F_\circ-mg)/(bv_\circ)=0.065$ and
$(F_\circ-mg)/(c{v_\circ}^2)=0.20$, respectively; $\chi^2$ is
smaller for the Poncelet model by a factor of two.  These fits
(not shown) are both acceptable, but not quite as nice as the one
displayed. Still, neither model predicts the observed penetration
depth scaling of Eq.~(\ref{junlaw}). The Bingham model gives
$d=(mv_\circ/b)[1-\alpha\ln(1+1/\alpha)]$ where
$\alpha=(F_\circ-mg)/(bv_\circ)$, and the Poncelet model gives
$d=[m/(2c)]\ln(1+1/\alpha)$ where
$\alpha=(F_\circ-mg)/(c{v_\circ}^2)$.  Furthermore the predicted
initial accelerations, and the final penetration depths, are
unphysical in the limit $v_\circ\rightarrow0$.

We now modify the Poncelet model along the lines suggested by
Tsimring and Volfson \cite{LevPG}.  Since the constant term
represents friction, it should depend on depth according to
hydrostatic pressure \cite{schiffer01,detlef3} and the geometry of
the granular medium near the projectile: $\Sigma F = -mg + [F(z) +
cv^2]$. For any $F(z)$ this can be recast as an ordinary
differential equation for kinetic energy vs position and solved by
use of an integrating factor, $\exp(-2cz/m)$.  To {\it exactly}
recover the observed penetration depth scaling,
Eq.~(\ref{junlaw}), and hence be consistent with the observed
depth-averaged stopping force, we find that the friction term must
be $F(z)/mg = [3(z / {d_\circ})^2 -1]\exp(-2|z|/d_1)+1$ where
$d_1=m/c$. This vanishes for shallow penetration and approaches a
constant for deep penetration, consistent with the special limits
advocated in Ref.~\cite{LevPG}. Altogether the projectile
acceleration in our model is
\begin{equation}
    a/g = [3(z/d_\circ)^2-1]\exp(-2|z|/d_1)+v^2/(gd_1).
\label{a:modponc}
\end{equation}
For high speeds and deep impacts the velocity term dominates. For
shallow impacts, the leading behavior is $a/g \approx -1 +
2|z|/d_1 + (3/d_\circ^2-2/d_1^2)z^2 + v^2/(gd_1)$.  For zero
impact speed, the initial acceleration is $-g$, as expected.

To compare our force model with data, it is convenient to work in
phase space since velocity vs depth can be computed directly as
\begin{equation}
    \left(v\over v_\circ\right)^2 =
    \left[1-{ |z/d|^3 - |z/d|(d_\circ/d)^2 \over 1 -
    (d_\circ/d)^2} \right] e^{-2|z|/d_1}.
\label{v:modponc}
\end{equation}
For $v_\circ=0$, the relation is $v^2=2gd_\circ(
|z/d_\circ|-|z/d_\circ|^3)\exp(-2|z|/d_1)$. Since the predicted
dynamics depend on the value of $d/d_\circ$, the apparent collapse
of data in Fig.~\ref{ZVAvsT} would then represent clustering about
an average to within measurement error. Setting $d/d_\circ=2.77$,
the average value for all our runs, the best one-parameter fit to
Eq.~(\ref{v:modponc}) is for $d/d_1=1.2$.  This gives an excellent
description of the data, as shown by the solid purple curves in
Figs.~\ref{ZVAvsT}-\ref{VAvsZ}.

As the ball comes to rest, in all acceptable fits, the time
becomes about $3d/v_\circ$ and the upward acceleration becomes one
to two $g$.  After stopping the acceleration abruptly vanishes, as
seen directly in measurement by accelerometer \cite{PBU}.  The
discontinuity is more apparent here in plots of acceleration vs
position or speed (Fig.~\ref{VAvsZ}) than in plots of acceleration
vs time (Fig.~\ref{ZVAvsT}c).


In conclusion, we have measured the dynamics of impact to further
constrain the form of the force law responsible for prior
observations of final penetration depth vs total drop distance,
$d\sim H^{1/3}$ \cite{jun, katie, mike1}. For two sphere sizes,
and a factor of two in drop distance, the position vs time data
appear to collapse when scaled by impact speed and final depth.
Several possible force laws can now be ruled out altogether.  Two
velocity-dependent force laws are consistent with dynamics but not
with penetration depth data.  The best candidate is a modification
of the Poncelet model along the lines suggested in
Ref.~\cite{LevPG}. It features inertial drag plus a particular
depth-dependent friction term designed to exactly recover the
penetration depth scaling of Eq.~(\ref{junlaw}).  The predicted
kinematics, Eqs.~(\ref{a:modponc}-\ref{v:modponc}), agree very
well with our new data.  The model has two important length
scales, the minimum penetration depth $d_\circ$ and an inertial
drag length $d_1$. The former is given by $d_\circ =
(0.14/\mu)^{3/2} (\rho_b/\rho_g)^{3/4} D_b$ where $\mu$ is the
tangent of the repose angle, $\rho_g$ is the grain density,
$\rho_b$ is the ball density, and $D_b$ is the ball diameter
\cite{mike1}.  For the system examined here, $d_1$ is about twice
as large as $d_\circ$ but we do not yet know the dependence of
$d_1$ on system properties. This could be deduced, and the model
could be further tested, by measurement of impact dynamics over a
broad range of conditions.  At one extreme, the ``dry quick sand''
examined in Ref.~\cite{detlef3} for $v_\circ=0$ is reproduced very
well by our model with $d_1=3d_\circ$.

\begin{acknowledgments}
We thank D. Lohse, L.S.~Tsimring, and P.B.~Umbanhowar for sharing
their results prior to publication. We thank T.C.~Lubensky and
P.T.~Korda for helpful discussions. This material is based upon
work supported by the National Science Foundation under grant
DMR-0305106.
\end{acknowledgments}

\bibliography{CraterRefs}

\begin{thebibliography}{20}
\expandafter\ifx\csname natexlab\endcsname\relax\def\natexlab#1{#1}\fi
\expandafter\ifx\csname bibnamefont\endcsname\relax
  \def\bibnamefont#1{#1}\fi
\expandafter\ifx\csname bibfnamefont\endcsname\relax
  \def\bibfnamefont#1{#1}\fi
\expandafter\ifx\csname citenamefont\endcsname\relax
  \def\citenamefont#1{#1}\fi
\expandafter\ifx\csname url\endcsname\relax
  \def\url#1{\texttt{#1}}\fi
\expandafter\ifx\csname urlprefix\endcsname\relax\def\urlprefix{URL }\fi
\providecommand{\bibinfo}[2]{#2}
\providecommand{\eprint}[2][]{\url{#2}}

\bibitem[{\citenamefont{Amato and Williams}(1998)}]{amato}
\bibinfo{author}{\bibfnamefont{J.~C.} \bibnamefont{Amato}} \bibnamefont{and}
  \bibinfo{author}{\bibfnamefont{R.~E.} \bibnamefont{Williams}},
  \bibinfo{journal}{Am. J. Phys.} \textbf{\bibinfo{volume}{66}},
  \bibinfo{pages}{141} (\bibinfo{year}{1998}).

\bibitem[{\citenamefont{Grasselli and Herrmann}(2001)}]{hans}
\bibinfo{author}{\bibfnamefont{Y.}~\bibnamefont{Grasselli}} \bibnamefont{and}
  \bibinfo{author}{\bibfnamefont{H.~J.} \bibnamefont{Herrmann}},
  \bibinfo{journal}{Gran. Matt.} \textbf{\bibinfo{volume}{3}},
  \bibinfo{pages}{201} (\bibinfo{year}{2001}).

\bibitem[{\citenamefont{Walsh et~al.}(2003)\citenamefont{Walsh, Holloway,
  Habdas, and de~Bruyn}}]{deBruyn}
\bibinfo{author}{\bibfnamefont{A.~M.} \bibnamefont{Walsh}},
  \bibinfo{author}{\bibfnamefont{K.~E.} \bibnamefont{Holloway}},
  \bibinfo{author}{\bibfnamefont{P.}~\bibnamefont{Habdas}}, \bibnamefont{and}
  \bibinfo{author}{\bibfnamefont{J.~R.} \bibnamefont{de~Bruyn}},
  \bibinfo{journal}{Phys. Rev. Lett.} \textbf{\bibinfo{volume}{91}},
  \bibinfo{pages}{104301} (\bibinfo{year}{2003}).

\bibitem[{\citenamefont{Thoroddsen and Shen}(2001)}]{siggi}
\bibinfo{author}{\bibfnamefont{S.~T.} \bibnamefont{Thoroddsen}}
  \bibnamefont{and} \bibinfo{author}{\bibfnamefont{A.~Q.} \bibnamefont{Shen}},
  \bibinfo{journal}{Phys. Fluids} \textbf{\bibinfo{volume}{13}},
  \bibinfo{pages}{4} (\bibinfo{year}{2001}).

\bibitem[{\citenamefont{Lohse et~al.}(2004{\natexlab{a}})\citenamefont{Lohse,
  Bergmann, Mikkelsen, Zeilstra, van~der Meer, Versluis, van~der Weele, van~der
  Hoef, and Kuipers}}]{detlef2}
\bibinfo{author}{\bibfnamefont{D.}~\bibnamefont{Lohse}},
  \bibinfo{author}{\bibfnamefont{R.}~\bibnamefont{Bergmann}},
  \bibinfo{author}{\bibfnamefont{R.}~\bibnamefont{Mikkelsen}},
  \bibinfo{author}{\bibfnamefont{C.}~\bibnamefont{Zeilstra}},
  \bibinfo{author}{\bibfnamefont{D.}~\bibnamefont{van~der Meer}},
  \bibinfo{author}{\bibfnamefont{M.}~\bibnamefont{Versluis}},
  \bibinfo{author}{\bibfnamefont{K.}~\bibnamefont{van~der Weele}},
  \bibinfo{author}{\bibfnamefont{M.}~\bibnamefont{van~der Hoef}},
  \bibnamefont{and} \bibinfo{author}{\bibfnamefont{H.}~\bibnamefont{Kuipers}},
  \bibinfo{journal}{Phys. Rev. Lett.} \textbf{\bibinfo{volume}{93}},
  \bibinfo{pages}{198003} (\bibinfo{year}{2004}{\natexlab{a}}).

\bibitem[{\citenamefont{Uehara et~al.}(2003)\citenamefont{Uehara, Ambroso,
  Ojha, and Durian}}]{jun}
\bibinfo{author}{\bibfnamefont{J.~S.} \bibnamefont{Uehara}},
  \bibinfo{author}{\bibfnamefont{M.~A.} \bibnamefont{Ambroso}},
  \bibinfo{author}{\bibfnamefont{R.~P.} \bibnamefont{Ojha}}, \bibnamefont{and}
  \bibinfo{author}{\bibfnamefont{D.~J.} \bibnamefont{Durian}},
  \bibinfo{journal}{Phys. Rev. Lett.} \textbf{\bibinfo{volume}{90}},
  \bibinfo{pages}{194301} (\bibinfo{year}{2003}), \bibinfo{note}{and erratum
  {\bf 91}, 149902 (2003)}.

\bibitem[{\citenamefont{Newhall and Durian}(2003)}]{katie}
\bibinfo{author}{\bibfnamefont{K.~A.} \bibnamefont{Newhall}} \bibnamefont{and}
  \bibinfo{author}{\bibfnamefont{D.~J.} \bibnamefont{Durian}},
  \bibinfo{journal}{Phys. Rev. E} \textbf{\bibinfo{volume}{68}},
  \bibinfo{pages}{060301R} (\bibinfo{year}{2003}).

\bibitem[{\citenamefont{de~Bruyn and Walsh}(2004)}]{deBruyn2}
\bibinfo{author}{\bibfnamefont{J.~R.} \bibnamefont{de~Bruyn}} \bibnamefont{and}
  \bibinfo{author}{\bibfnamefont{A.~M.} \bibnamefont{Walsh}},
  \bibinfo{journal}{Can. J. Phys.} \textbf{\bibinfo{volume}{82}},
  \bibinfo{pages}{439} (\bibinfo{year}{2004}).

\bibitem[{\citenamefont{Zheng et~al.}(2004)\citenamefont{Zheng, Wang, and
  Qiu}}]{zheng}
\bibinfo{author}{\bibfnamefont{X.-J.} \bibnamefont{Zheng}},
  \bibinfo{author}{\bibfnamefont{Z.-T.} \bibnamefont{Wang}}, \bibnamefont{and}
  \bibinfo{author}{\bibfnamefont{Z.-G.} \bibnamefont{Qiu}},
  \bibinfo{journal}{European Physical Journal E} \textbf{\bibinfo{volume}{13}},
  \bibinfo{pages}{321} (\bibinfo{year}{2004}).

\bibitem[{\citenamefont{Ambroso et~al.}(2004)\citenamefont{Ambroso, Santore,
  Abate, and Durian}}]{mike1}
\bibinfo{author}{\bibfnamefont{M.~A.} \bibnamefont{Ambroso}},
  \bibinfo{author}{\bibfnamefont{C.~R.} \bibnamefont{Santore}},
  \bibinfo{author}{\bibfnamefont{A.~R.} \bibnamefont{Abate}}, \bibnamefont{and}
  \bibinfo{author}{\bibfnamefont{D.~J.} \bibnamefont{Durian}}
  (\bibinfo{year}{2004}), \bibinfo{note}{cond-mat/0411231}.

\bibitem[{\citenamefont{Brown and Richards}(1970)}]{brown}
\bibinfo{author}{\bibfnamefont{R.~L.} \bibnamefont{Brown}} \bibnamefont{and}
  \bibinfo{author}{\bibfnamefont{J.~C.} \bibnamefont{Richards}},
  \emph{\bibinfo{title}{Principles of Powder Mechanics}}
  (\bibinfo{publisher}{Pergamon Press}, \bibinfo{address}{Oxford},
  \bibinfo{year}{1970}).

\bibitem[{\citenamefont{Jaeger et~al.}(1996)\citenamefont{Jaeger, Nagel, and
  Behringer}}]{jnb}
\bibinfo{author}{\bibfnamefont{H.~M.} \bibnamefont{Jaeger}},
  \bibinfo{author}{\bibfnamefont{S.~R.} \bibnamefont{Nagel}}, \bibnamefont{and}
  \bibinfo{author}{\bibfnamefont{R.~P.} \bibnamefont{Behringer}},
  \bibinfo{journal}{Rev. Mod. Phys.} \textbf{\bibinfo{volume}{68}},
  \bibinfo{pages}{1259} (\bibinfo{year}{1996}).

\bibitem[{\citenamefont{Duran}(2000)}]{duran}
\bibinfo{author}{\bibfnamefont{J.}~\bibnamefont{Duran}},
  \emph{\bibinfo{title}{Sands, powders, and grains: An introduction to the
  physics of granular materials}} (\bibinfo{publisher}{Springer},
  \bibinfo{address}{NY}, \bibinfo{year}{2000}).

\bibitem[{\citenamefont{Ciamarra et~al.}(2004)\citenamefont{Ciamarra, Lara,
  Lee, Goldman, Vishik, and Swinney}}]{ciamarra}
\bibinfo{author}{\bibfnamefont{M.~P.} \bibnamefont{Ciamarra}},
  \bibinfo{author}{\bibfnamefont{A.~H.} \bibnamefont{Lara}},
  \bibinfo{author}{\bibfnamefont{A.~T.} \bibnamefont{Lee}},
  \bibinfo{author}{\bibfnamefont{D.~I.} \bibnamefont{Goldman}},
  \bibinfo{author}{\bibfnamefont{I.}~\bibnamefont{Vishik}}, \bibnamefont{and}
  \bibinfo{author}{\bibfnamefont{H.~L.} \bibnamefont{Swinney}},
  \bibinfo{journal}{Phys. Rev. Lett.} \textbf{\bibinfo{volume}{92}},
  \bibinfo{pages}{194301} (\bibinfo{year}{2004}).

\bibitem[{\citenamefont{Lohse et~al.}(2004{\natexlab{b}})\citenamefont{Lohse,
  Rauhe, Bergmann, and van~der Meer}}]{detlef3}
\bibinfo{author}{\bibfnamefont{D.}~\bibnamefont{Lohse}},
  \bibinfo{author}{\bibfnamefont{R.}~\bibnamefont{Rauhe}},
  \bibinfo{author}{\bibfnamefont{R.}~\bibnamefont{Bergmann}}, \bibnamefont{and}
  \bibinfo{author}{\bibfnamefont{D.}~\bibnamefont{van~der Meer}},
  \bibinfo{journal}{Nature} \textbf{\bibinfo{volume}{432}},
  \bibinfo{pages}{689} (\bibinfo{year}{2004}{\natexlab{b}}).

\bibitem[{\citenamefont{Daniels et~al.}(2004)\citenamefont{Daniels, Coppock,
  and Behringer}}]{RPB04}
\bibinfo{author}{\bibfnamefont{K.~E.} \bibnamefont{Daniels}},
  \bibinfo{author}{\bibfnamefont{J.~E.} \bibnamefont{Coppock}},
  \bibnamefont{and} \bibinfo{author}{\bibfnamefont{R.~P.}
  \bibnamefont{Behringer}}, \bibinfo{journal}{Chaos}
  \textbf{\bibinfo{volume}{14}}, \bibinfo{pages}{S4} (\bibinfo{year}{2004}).

\bibitem[{\citenamefont{Umbanhowar}(2004)}]{PBU}
\bibinfo{author}{\bibfnamefont{P.~B.} \bibnamefont{Umbanhowar}}
  (\bibinfo{year}{2004}), \bibinfo{note}{private communication}.

\bibitem[{\citenamefont{Backman and Goldsmith}(1978)}]{backman78}
\bibinfo{author}{\bibfnamefont{M.~E.} \bibnamefont{Backman}} \bibnamefont{and}
  \bibinfo{author}{\bibfnamefont{W.}~\bibnamefont{Goldsmith}},
  \bibinfo{journal}{International Journal of Engineering Science}
  \textbf{\bibinfo{volume}{16}}, \bibinfo{pages}{1} (\bibinfo{year}{1978}).

\bibitem[{\citenamefont{Tsimring and Volfson}(2005)}]{LevPG}
\bibinfo{author}{\bibfnamefont{L.~S.} \bibnamefont{Tsimring}} \bibnamefont{and}
  \bibinfo{author}{\bibfnamefont{D.}~\bibnamefont{Volfson}},
  \bibinfo{journal}{preprint}  (\bibinfo{year}{2005}), \bibinfo{note}{submitted
  to Powders and Grains 2005}.

\bibitem[{\citenamefont{Albert et~al.}(2001)\citenamefont{Albert, Sample,
  Morss, Rajagopalan, Barabasi, and Schiffer}}]{schiffer01}
\bibinfo{author}{\bibfnamefont{I.}~\bibnamefont{Albert}},
  \bibinfo{author}{\bibfnamefont{J.~G.} \bibnamefont{Sample}},
  \bibinfo{author}{\bibfnamefont{A.~J.} \bibnamefont{Morss}},
  \bibinfo{author}{\bibfnamefont{S.}~\bibnamefont{Rajagopalan}},
  \bibinfo{author}{\bibfnamefont{A.~L.} \bibnamefont{Barabasi}},
  \bibnamefont{and} \bibinfo{author}{\bibfnamefont{P.}~\bibnamefont{Schiffer}},
  \bibinfo{journal}{Phys. Rev. E} \textbf{\bibinfo{volume}{64}},
  \bibinfo{pages}{061303} (\bibinfo{year}{2001}), \bibinfo{note}{and references
  therein}.

\end{thebibliography}

\end{document}